\documentclass[9pt,twocolumn,twoside]{opticajnl}
\journal{opticajournal} 
\usepackage[utf8]{inputenc}
\DeclareUnicodeCharacter{03BB}{$\lambda$}
\DeclareUnicodeCharacter{03C0}{$\pi$}
\setboolean{shortarticle}{false}

\usepackage{subcaption}
\makeatletter
\renewcommand*\subcaption@label{\caption@withoptargs\subcaption@@label}
\makeatother
\usepackage{stfloats}
\usepackage{placeins}
\usepackage{multirow}
\usepackage{siunitx}
\usepackage{physics}
\DeclareSIUnit\baud{Baud}

\usepackage{lineno}

\usepackage{todonotes}
\definecolor{edo}{RGB}{0, 150, 0}

\definecolor{mattia}{HTML}{FF8000}

\usepackage[dvipsnames]{xcolor}
\usepackage[normalem]{ulem}

\title{ Squeezed- and  coherent-state quantum key distribution over a deployed hybrid fibre-free-space channel}

\author[1, *]{Adnan A.E. Hajomer}
\author[1]{Huy Q. Nguyen}
\author[2]{Ivan Derkach}
\author[3]{Andreas B. Kidmose}
\author[4]{Edoardo  Rossi}
\author[4]{Mattia Sabatini}
\author[4]{Yoann  Pietri}
\author[4]{Marco Avesani}
\author[4]{Francesco Vedovato}
\author[5]{Michael Hentschel}
\author[2]{Radim Filip}
\author[4]{Giuseppe Vallone}
\author[2]{Vladyslav Usenko}
\author[1,*]{Tobias Gehring}
\author[3]{Søren Forchhammer}
\author[4]{Paolo  Villoresi}
\author[1,*]{Ulrik L. Andersen}

\affil[1]{Center for Macroscopic Quantum States (bigQ), Department of Physics, Technical University of Denmark, 2800 Kongens Lyngby, Denmark}
\affil[2]{Department of Optics, Faculty of Science, Palacky University, 17. listopadu 12, 771 46 Olomouc, Czech Republic}
\affil[3]{Department of Electrical and Photonics Engineering
Technical University of Denmark, 2800 Kongens Lyngby, Denmark}
\affil[4]{Dipartimento di Ingegneria dell’Informazione, Università degli Studi di Padova, via Gradenigo 6B, IT-35131 Padova, Italy}
\affil[5]{AIT  Austrian  Institute  of  Technology,  Center  for  Digital  Safety\&Security,  Giefinggasse  4,  1210  Vienna, Austria}
\affil[*]{Corresponding authors: aaeha@dtu.dk, tobias.gehring@fysik.dtu.dk, ulrik.andersen@fysik.dtu.dk}

\begin{abstract}
Quantum networks will combine optical fibre with free-space links, yet
continuous-variable quantum key distribution (CV-QKD) has been developed 
predominantly for one medium or the other, while operation across
concatenated fibre--free-space channels remains largely unexplored.
The two media impose contrasting requirements: fibre transmission is stable and
permits long processing intervals, whereas atmospheric propagation imposes
transmittance fluctuations that degrade security and must be resolved on short
timescales. Here we demonstrate a locally generated local oscillator CV-QKD with both
Gaussian-modulated coherent and squeezed states over a deployed hybrid channel
comprising a 620-m free-space link and 2~km of deployed fibre,
with a total loss up to 20~dB. Rather than adapting the optics to each medium, we move channel adaptation to the post-processing, through a unified adaptive post-processing framework coupling transmittance-based clustering, residual-fading
mitigation by covariance-matrix averaging or de-fading, and rate-adaptive blind
reconciliation, which alone recovers up to 19\% additional key. The same adaptive-processing principle is applied to both protocols, while accounting for their different security analyses and statistical requirements, yielding asymptotic secret-key rates of 0.42 Mbit/s for the coherent-state protocol and 0.93 Mbit/s for the squeezed-state protocol under the respective channel conditions, and
establishing squeezed-state CV-QKD over a deployed atmospheric channel. These results show that adaptation to the transmission medium can largely be transferred to the data-processing layer, providing a route towards heterogeneous quantum
networks spanning fibre, terrestrial free-space and satellite links.
\end{abstract}

\setboolean{displaycopyright}{false} 

\begin{document}
  
\maketitle

\section{Introduction}

Quantum networks will need to operate across multiple transmission
infrastructures. Optical fibre provides stable, low-loss connectivity for
terrestrial networks, whereas free-space links enable connectivity where fibre
deployment is impractical and offer a route towards mobile and satellite
quantum communication~\cite{pirandola2020advances}. Practical networks will
therefore be heterogeneous, combining both media within a single
infrastructure, and will require quantum communication systems capable of
operating across the resulting concatenated channels. Continuous-variable
quantum key distribution (CV-QKD) is a natural candidate for this role: by
encoding information in the quadratures of the electromagnetic field and
recovering it with coherent detection, CV-QKD can reuse standard
telecommunication components and infrastructure~\cite{usenko2026review}. Over
fibre the approach is mature, with demonstrations of high secret-key
rates~\cite{hajomer2024continuous,hajomer2026chip,ng2026gigabit}, transmission
beyond 100~km~\cite{zhang2020long}, coexistence with classical
channels~\cite{hajomer2025coexistence}, and multi-user quantum access
networks~\cite{hajomer2024continuous,pan2025high}. 

Free-space CV-QKD has advanced considerably in parallel, with secret-key
distribution reported over both deployed and emulated atmospheric
channels~\cite{zhan2026long,zheng2025free,yin2025all,jaksch2026composable,liao2025high}.
Two features of this body of work are relevant here. First, deployed
demonstrations have relied predominantly on a transmitted local oscillator, in
which the reference beam propagates through the untrusted
channel~\cite{zhan2026long,zheng2025free,yin2025all}. This increases system
complexity, introduces known security vulnerabilities, and is poorly suited to
a path that alternates between fibre and free space, where the reference must
survive both media and their differing mode and polarisation behaviour.
Locally generated local oscillator (LLO) schemes remove this requirement and
have recently been applied to free-space CV-QKD, but so far only under
emulated channels~\cite{jaksch2026composable,liao2025high}. Second, CV-QKD systems have been developed for either fibre or free-space
transmission. Operation across a channel that concatenates the two remains
largely unexplored, although such links are precisely what heterogeneous
networks require.

The difficulty is that the two media impose opposing requirements on the
processing of the measured data. Fibre transmission is stable, and its channel
parameters can be estimated over long intervals. Atmospheric propagation,
together with the time-dependent coupling of the received field into
single-mode fibre, instead causes the transmittance to vary during
acquisition, so that the channel is described by a distribution rather than by
a single value. Such fading does not merely add loss: it degrades CV-QKD
security, and suppressing that penalty requires resolving the channel on
timescales shorter than its variation~\cite{usenko2012entanglement}.
Strategies for doing so exist --- transmittance-based
binning~\cite{ruppert2019fading,hosseinidehaj2021composable}, active
compensation~\cite{pirandola2021limits}, security analyses that retain the
residual fluctuations~\cite{derkach2020squeezing,oruganti2025continuous}, and
adaptive information
reconciliation~\cite{DBLP:journals/qic/Martinez-MateoEM12} --- but have
largely been studied in isolation, although they are coupled through the same
channel statistics: how the transmittance distribution is partitioned sets the
residual fading each portion of the data must tolerate, and simultaneously
determines the statistics available for parameter estimation and the
signal-to-noise ratio that reconciliation must match. How to combine them for
a channel that is stable along one segment and fluctuating along another has
not been addressed.

Squeezed states sharpen this problem, and at the same time supply a reason to
solve it. Reducing the quadrature noise below the vacuum level improves the
tolerance of the protocol to excess noise \cite{Madsen2012} and to imperfect
reconciliation \cite{Usenko2011}, so that
squeezed-state protocols relax the noise requirement under which a positive key can be extracted \cite{nguyen2025practical}. Squeezing then becomes particularly helpful in fading channels \cite{derkach2020squeezing} provided that squeezing level and anti-squeezing noise are controlled \cite{oruganti2025continuous}. This advantage, however, has so far been demonstrated only in laboratory and fibre-based settings \cite{nguyen2025practical}, as squeezing is degraded by optical loss and further compromised by transmittance fluctuations. To our knowledge, squeezed-state CV-QKD has not previously been demonstrated over a deployed atmospheric channel. It also
carries a specific statistical cost: estimating the squeezing and
anti-squeezing variances required by the security analysis demands longer processing
frames than the coherent-state protocol. Fading
mitigation pushes the frame length down; squeezed-state parameter estimation
pushes it up. Whether squeezing remains a usable resource over a deployed
atmospheric channel therefore depends on whether this conflict of timescales
can be resolved.

Here we demonstrate LLO CV-QKD with both Gaussian-modulated coherent and
squeezed states over a deployed hybrid quantum channel comprising a 620-m
free-space link and 2~km of fibre, with a total loss of
15 to 20~dB. Within each protocol, the optical architecture does not require reconfiguration when operating across the fibre and free-space segments, while the same adaptive post-processing principle applies to both coherent- and squeezed-state protocols. Our adaptive post-processing framework combines transmittance-based clustering, residual-fading mitigation through covariance-matrix averaging or de-fading, and rate-adaptive blind reconciliation. Although the same framework is used for both protocols, its implementation accounts for differences in their security analysis, the parameters that must be estimated, and the processing-frame length, which determines the resolution with which the transmittance distribution can be partitioned. The framework enables secret-key
extraction across the range of channel conditions encountered on the deployed
link, yielding asymptotic secret-key rates of \textcolor{black}{0.42~Mbit/s} for the
coherent-state protocol and \textcolor{black}{0.93~Mbit/s} for the squeezed-state
protocol. These results establish CV-QKD operation across concatenated
fibre and free-space infrastructure, show that squeezing remains available as
a resource outside the laboratory, and indicate that adaptation to the
transmission medium can largely be transferred from the optics to the
data-processing layer.

\section{CV-QKD system and hybrid channel}\label{sec:experiment}

\begin{figure}[t]
    \centering
    \includegraphics[width=0.99\linewidth]{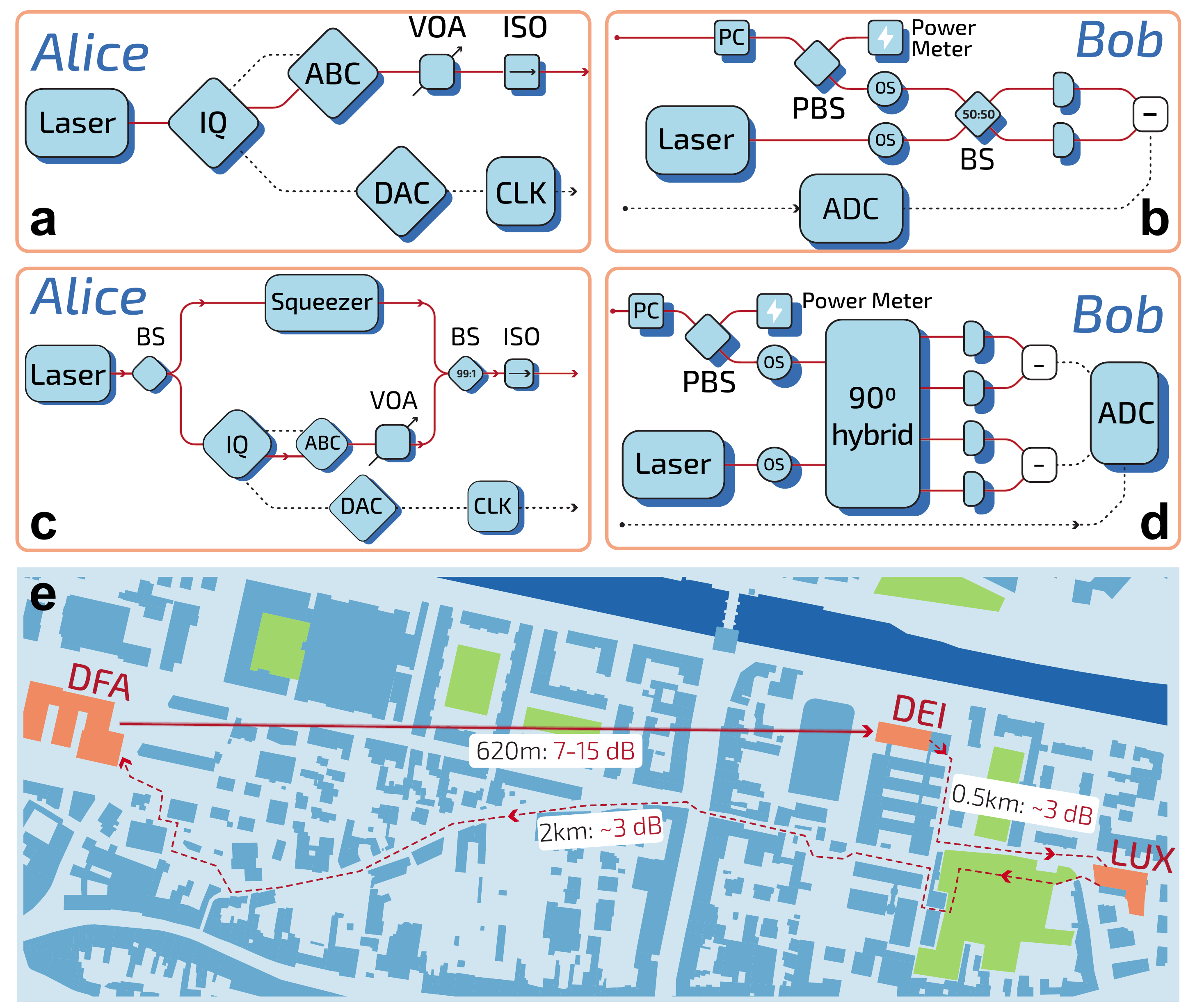}
    \caption{
Experimental configuration of the hybrid fiber--free-space CV-QKD system.
\textbf{a}, Coherent-state transmitter.
\textbf{b}, Heterodyne receiver with a locally generated local oscillator.
\textbf{c}, Squeezed-state transmitter.
\textbf{d}, Phase-diverse receiver based on a 90$^\circ$ optical hybrid.
\textbf{e}, Hybrid quantum link deployed in Padova, connecting LUX, DFA, and
DEI through fiber and free-space segments. The indicated values denote the
approximate losses of the corresponding link segments.
IQ, in-phase and quadrature modulator; ABC, automatic bias controller;
VOA, variable optical attenuator; ISO, optical isolator; DAC, digital-to-analog converter; CLK, reference clock;
PC, polarization controller; OS, optical switch; PBS, polarization beam
splitter; BS, beam splitter; ADC, analog-to-digital converter.} 
    \label{fig:network}
\end{figure}

Figure~\ref{fig:network} presents the CV-QKD system and the hybrid
fibre--free-space testbed deployed in Padova, Italy~\cite{piccia_DFA_DEI, Bolanos:26}. The system implements both
the Gaussian-modulated coherent- and squeezed-state protocols using a 
 LLO scheme. The two protocols share the same modulation,
synchronisation and digital signal processing (DSP), and differ in the
quantum-state source, the detection front end, and the security analysis
applied to the recovered data. Operating both over the same deployed link
therefore exercises the adaptive post-processing framework on two protocols
with different statistical requirements, using a common transmitter and DSP
chain. The same system is operated over two
channel configurations, described in the following subsection, which expose it
to different end-to-end loss and fluctuation regimes.

\subsection{Coherent-state CV-QKD system}

Alice employs a continuous-wave laser at 1550~nm as the optical carrier (Fig.~\ref{fig:network}a). Gaussian-distributed symbols generated from a quantum random number generator are pulse-shaped and applied to an IQ modulator operated in optical single-sideband suppressed-carrier mode, thereby encoding information in the amplitude and phase quadratures. The symbols are generated at 125~MBaud, upsampled to 1~GSample/s, and shaped using a root-raised-cosine filter with roll-off 0.2. To avoid low-frequency technical noise, the quantum signal is shifted to 120~MHz, while a strong pilot tone at 220~MHz is frequency-division multiplexed with the signal for frequency and phase recovery. The modulator operating point is stabilized using an automatic bias controller, and the modulation variance is adjusted using a variable optical attenuator. An optical isolator at the transmitter output suppresses back-reflections and mitigates Trojan-horse attacks.

At Bob (Fig.~\ref{fig:network}b), the quantum signal is detected using RF heterodyne detection with a local oscillator generated by an independent continuous-wave laser. The local oscillator is detuned from Alice's laser by approximately 260~MHz. Polarization overlap between the quantum signal and local oscillator is optimized using a polarization controller and monitored through a polarization beam splitter and power meter. The detected signal is acquired using a balanced detector with a shot-noise-limited bandwidth of 250~MHz and digitized at 1~GSample/s. To synchronize Alice and Bob, optical clock and trigger signals at 1310~nm and 1553.33~nm, respectively, are wavelength-multiplexed and distributed to Bob over optical fiber. The quantum symbols are recovered through a DSP chain comprising whitening filtering, frequency and phase estimation using an unscented Kalman filter, timing synchronization, and matched filtering~\cite{hajomer2024long}. The detector efficiency of $0.68$ and electronic noise are calibrated independently and treated as trusted
(Sec.~\ref{sec:protocols}).

\subsection{Squeezed-state CV-QKD system}
The squeezed-state system retains the same modulation, synchronization, and DSP, but replaces the coherent-state source and receiver front end. At Alice (Fig.~\ref{fig:network}c), squeezed light is generated using a cascaded pair of periodically poled lithium niobate waveguides. The first waveguide performs second-harmonic generation (SHG), converting 1550~nm light to 775~nm. The second waveguide is pumped by the 775~nm field to generate squeezed light through parametric down-conversion. Residual pump light is removed using a wavelength-division multiplexer, and the squeezed state is displaced by interferometrically overlapping it with  coherent states.

Because the squeezing bandwidth is on the order of terahertz and therefore exceeds the detector bandwidth, RF heterodyne detection is not used. Instead, Bob employs intradyne detection with a 90° optical hybrid (Fig.~\ref{fig:network}d), providing simultaneous access to both field quadratures. The additional 1 dB insertion loss introduced by the optical hybrid results in an overall detection efficiency of 0.54. The subsequent DSP follows that of the coherent-state system, with additional squeezing-angle estimation and quadrature remapping~\cite{nguyen2025practical}.

Access to both quadratures allows the squeezing variance $V_S$ and the
anti-squeezing noise $V_{AS}$ required by the security analysis
(Sec.~\ref{sec:protocols}) to be estimated alongside the channel parameters.
Their estimation is statistically more demanding than that of the
coherent-state protocol and sets the longer processing frames used throughout
the analysis of the squeezed-state data (Sec.4~\ref{sec:pipeline}).

\subsection{Hybrid channel testbed}
\label{sec:freespace}

\begin{figure}
    \centering
    \includegraphics[width=0.99\linewidth]{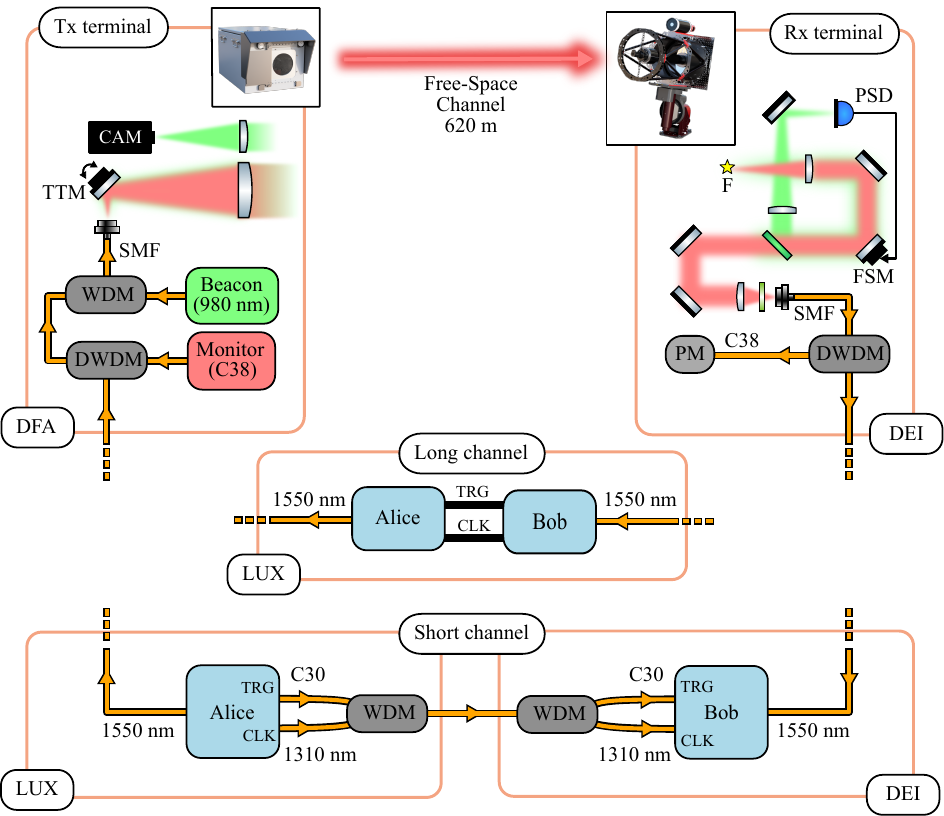}
    \caption{
Free-space optical terminals and signal routing for the two hybrid-channel configurations. The upper panels show the transmitter (Tx) at DFA and receiver (Rx) at DEI, connected by the 620-m free-space link; the lower panels show the long- and short-channel configurations. DWDM, dense wavelength-division multiplexer; WDM, wavelength-division multiplexer; SMF, single-mode fiber; TTM, tip--tilt mirror; CAM, camera; F, receiver focal plane; FSM, fast-steering mirror; PSD, position-sensitive detector; PM, power meter; TRG, trigger; CLK, clock.}
    \label{fig:free-space}
\end{figure}

The hybrid channel was deployed across three sites at the University of
Padova: the Luxor Laboratory (LUX), the Department of Physics and Astronomy
(DFA), and the Department of Information Engineering (DEI)
(Fig.~\ref{fig:network}e and Fig.~\ref{fig:free-space}). The quantum signal
travels from Alice at LUX to the free-space transmitter at DFA over deployed
fibre, introducing approximately 3~dB of loss, and then propagates over a
620-m atmospheric link to a sub-meter aperture telescope at DEI, where it is
coupled into single-mode fibre. The loss of the fibre segment is
dominated by connectors and patch panels rather than by fibre attenuation. The
loss of the free-space segment is set by the atmospheric conditions and by the
resulting coupling efficiency into single-mode fibre, and ranges from
approximately 7 to 15~dB across the conditions encountered during the
measurement campaign.

We investigate two configurations. In the \textit{short-channel}
configuration (LUX--DFA--DEI), Bob is located at DEI and receives the signal
directly after the free-space link, giving a total loss of approximately 10 to
18~dB. In the \textit{long-channel} configuration (LUX--DFA--DEI--LUX), the
signal is routed back to Bob at LUX through an additional deployed fibre span
of approximately 3~dB loss, resulting in 2~km of fibre in the complete
link and a total loss of approximately 13 to 21~dB. The long-channel
configuration is of
particular relevance for deployed infrastructure: Alice and Bob are co-located
at a single node, and the atmospheric hop is an intermediate span rather than
the endpoint of the link. This is the topology in which a free-space segment
would be used to bridge a gap in terrestrial fibre, and it requires the
receiver to operate on a signal that has traversed both media and been
re-coupled into fibre.

The free-space link consists of an optical transmitter (Tx) installed at DFA
and a receiver (Rx) integrated into an optical ground station at DEI, both
described in detail in
Ref.~\cite{Bolanos:26}. At the transmitter,
the quantum signal arriving from LUX is multiplexed with a beacon and a
channel-monitoring signal before free-space transmission. At DEI, the optical
ground station collects the incoming beams and actively stabilises their
coupling into single-mode fibre: the beacon is detected by a
position-sensitive detector, which provides feedback to a fast-steering mirror
that compensates beam displacement in the receiver focal plane and improves
fibre coupling~\cite{Bolanos:26}. After fibre
coupling, a 100~GHz dense wavelength-division multiplexer centered at C38
(1546.92~nm) separates the channel-monitoring signal from the quantum signal.

The channel-monitoring signal is generated by a laser at 1546.92~nm and
recorded at 10~Hz, providing a continuous measure of the transmission of the
free-space segment independently of the quantum signal. It is used to
characterise the atmospheric link during each acquisition. In addition, atmospheric conditions were recorded in parallel by the weather station at
DEI. The two channel configurations expose the same CV-QKD
system to different transmittance distributions, which are analyzed
in Sec.~\ref{sec:adaptive-processing}.

\section{CV-QKD protocols}
\label{sec:protocols}

\subsection{Coherent-state protocol}

We evaluate the coherent-state CV-QKD system using the Gaussian-modulated no-switching protocol \cite{weedbrook2004quantum} under the trusted-detector assumption, where the detector inefficiency and electronic noise are calibrated and not attributed to Eve \cite{usenko2016trusted}. The lower bound on the asymptotic secret-key rate (SKR), expressed in bits per channel use, is

\begin{equation}
\label{eq:coh-skr}
R^C=
\max\left\{
0,\,
\beta I(A:B)^{x,p}
-S(AB^xB^p)
+S(A|B^xB^p)
\right\}.
\end{equation}
Here, $I(A:B)^{x,p}$ is the total mutual information between mode of Alice $A$ and mode of Bob $B$ in the $x$ and $p$ quadratures, and $\beta\in[0,1]$ is the reconciliation efficiency. Following the purification-based analysis of security against optimal Gaussian collective attacks \cite{Navascues2006, Garcia2006}, Eve's accessible information is bounded by the Holevo quantity
$\chi_E=S(AB^xB^p)-S(A|B^xB^p)$, where $B^{x(p)}$ denotes the mode received by Bob whose state was measured in the $x(p)$ quadrature. 
A positive secret key is therefore obtained when the reconciled mutual information exceeds the corresponding Holevo bound.

The key rate depends on the modulation variance $V_M$, detector efficiency $\eta_D$, electronic noise $\nu_{\mathrm{el}}$, channel transmittance $\eta$, and excess noise $\varepsilon$. Channel properties $\eta$ and $\varepsilon$ are estimated and bounded from the measured data during post-processing. Detector parameters are calibrated and monitored, thus allowing to treat them as trusted devices in the security analysis. Further details of the security analysis, system optimization, and parameter estimation are given in Refs.~\cite{leverrier2015composable,hajomer2024long,hajomer2025coexistence}.

\subsection{Squeezed-state protocol}

In the squeezed-state protocol, information is encoded on a displaced squeezed
state, and Bob's measurement provides access to both
quadratures~\cite{derkach2020squeezing,oruganti2025continuous}. The
anti-squeezed-quadrature outcomes are publicly disclosed and incorporated into
the security analysis, where they serve to characterise source imperfections
and channel-induced noise~\cite{nguyen2025digital,nguyen2025practical,%
oruganti2025continuous}. The corresponding asymptotic secret-key rate is

\begin{equation}
\label{eq:sq-skr}
R^S=
\max\left\{
0,\,
\beta I(A:B)^x
-S(AB^x|B^p)
+S(A|B^xB^p)
\right\},
\end{equation}
in which entropy conditioning on Bob's measurement of $p$-quadrature $B^p$ indicates the public disclosure of the
anti-squeezed quadrature.

Beyond the parameters required for the coherent-state protocol, this analysis
requires the squeezing variance $V_S$ and the additional noise $V_{AS}$ in the
anti-squeezed quadrature. Both are estimated from the measured data rather
than assumed, and the precision of that estimation is what sets the processing
frame length used throughout Sec.~\ref{sec:adaptive-processing}. The detectors
are again treated as trusted. Further details of the protocol and its security
analysis are given in Ref.~\cite{oruganti2025continuous}.

In the following, these bounds are evaluated under the time-varying
transmittance of the deployed hybrid channel. For the fluctuating channel considered here, $\eta$ and $\varepsilon$ are
additionally estimated for each processing frame; these per-frame estimates form the basis of the adaptive processing described in Sec.~\ref{sec:adaptive-processing}.

\section{Adaptive processing of fluctuating channels}
\label{sec:adaptive-processing}

The hybrid channel combines comparatively stable fibre transmission with a
fluctuating free-space segment. In the latter, atmospheric propagation and
time-dependent coupling into single-mode fibre cause the channel transmittance
to vary during data acquisition, an effect known as fading. The channel is
then characterized not by a single transmittance $\eta$ but by a probability
distribution of transmittance (PDT), $\mathcal{P}(\eta)$. 

For a fixed-transmittance channel the received Gaussian-modulated state
remains Gaussian. In a fading channel, measurements acquired at different
transmittances form a statistical mixture of Gaussian states that is generally
non-Gaussian \cite{dong2010continuous}. Security can nevertheless be bounded using Gaussian extremality applied to the covariance matrix of the resulting ensemble, at the cost of an additional fading-induced noise
penalty ~\cite{usenko2012entanglement}. This penalty is 
proportional to variance-dependent excess noise $\varepsilon_x^f=\operatorname{Var}(\sqrt{\eta})\left(V_S+V_{M}-1\right)$ and $\varepsilon_p^f=\operatorname{Var}(\sqrt{\eta})\left(1 / V_S+V_{M}+V_{AS}-1\right)$ in $x$ - and $p$-quadratures respectively in the channel with fixed but effective and lower transmittance $\langle\sqrt{\eta}\rangle^2$ ~\cite{oruganti2025continuous}. The noise is determined by the fluctuations of the square root of the channel transmittance, $Var(\sqrt{\eta}) = \langle\eta\rangle-\langle\sqrt{\eta}\rangle^{2}$,
and increases with both $\mathrm{Var}(\sqrt{\eta})$ and the modulated state variance  \cite{usenko2012entanglement,derkach2020squeezing}. We refer to this method of incorporating fading into the security analysis as \textit{covariance-matrix averaging} (CMA). Consequently, channels with similar mean transmittance can yield substantially different secret-key rates depending on the strength of their transmittance fluctuations. Improving the key rate under CMA therefore requires not only a higher mean transmittance but also suppressed fading within the underlying transmittance distribution — both of which can be achieved through data post-selection (see Sec.4\ref{sec:threshold-and-binning}). 

An alternative to CMA is \textit{de-fading} \cite{pirandola2021limits}, which suppresses the fluctuations directly rather than folding them into the security analysis. Each measurement, associated with an estimated transmittance $\eta_j$, is attenuated to a predetermined reference value $\eta_{th}\leq\eta_j$, corresponding to the mapping $\eta_j\to\eta_{th}$, i.e., an effective pure-loss channel with transmittance $\eta_{\mathrm{th}}/\eta_j$ added after the receiver's purification of electronic noise. This differs from the coarser treatment used for fast-fading channels, where the lowest transmittance among a group of measurements is simply assigned to all of them \cite{papanastasiou2018continuous,castilloCeleita2026a}. The de-fading also scales the excess noise in the channel, but compensates it with additional Gaussian noise. This downshifting therefore eliminates the transmittance fluctuations, but lowers the overall signal-to-noise ratio (SNR) of the data.\\

\subsection{Conventional post-selection and binning}
\label{sec:threshold-and-binning}

A direct approach to reducing the fading penalty $Var(\sqrt{\eta})$ is threshold post-selection, in which measurements below a transmittance threshold $\eta_{\mathrm{th}}$ are discarded and only those with $\eta\geq\eta_{\mathrm{th}}$ are retained for key extraction \cite{pirandola2021limits,yao2025continuous,li2026security}. Increasing $\eta_{\mathrm{th}}$ generally reduces the residual transmittance fluctuations, but simultaneously decreases the fraction of data available for key extraction. The threshold must therefore balance the secret-key rate of the retained data against the fraction of measurements that remain. Its weighted secret-key contribution can be expressed as

\begin{equation}
R^{\eta_{\mathrm{th}}}
=
p(\eta_{\mathrm{th}})
r\!\left[
\mathrm{Var}(\sqrt{\eta}),
\langle\eta\rangle,
\langle\varepsilon\rangle
\right],
\label{eq:threshold}
\end{equation}
where $p(\eta_{\mathrm{th}})$ denotes the retained fraction of the measurement.

An alternative is to discretize the continuous PDT by partitioning the transmittance range into uniform bins of width $\Delta\eta$, where each bin defines contains events with $\eta\in\left[\eta_i,\eta_i+\Delta\eta\right)$ which significantly reduced loss fluctuations within that interval \cite{usenko2018stabilization,ruppert2019fading,hosseinidehaj2021composable}. The overall SKR is obtained by summing the per-bin key rates, weighted by the fraction of data contained in each bin:

\begin{equation}
R^\Sigma =
\sum_{j=1}^{K}p_j\max\{0,r_j\},
\label{eq:weightedSKR}
\end{equation}
where $p_j$ is the fraction of data assigned to the $j$th bin and $r_j$ is its corresponding secret-key rate. Narrow bins reduce the residual fluctuations $\mathrm{Var}(\sqrt{\eta_j})$, but contain fewer measurements, whereas wider bins retain more data at the cost of a larger fading penalty. Conventional binning therefore introduces a trade-off between suppressing fading and retaining sufficient data for reliable parameter estimation and key extraction.

For both uniform binning and threshold post-selection, choosing the bin width $\Delta\eta$ or the threshold $\eta_{th}$ is not straightforward, since the PDT alone does not capture excess noise variations across the data. Consequently, a choice that looks good from transmittance statistics alone may still yield a suboptimal SKR once noise is accounted for. In practice, determining $\Delta\eta$ or $\eta_{th}$ requires an optimization step that incorporates the noise behavior, which typically becomes clear only after error correction and subsequent parameter estimation on the candidate partitions.\\
Both approaches are already used in free-space discrete-variable QKD, particularly satellite links, to control the quantum bit error rate by restricting key extraction to the most favorable part of a pass or by segmenting it into separate keys \cite{sidhu2023finite, marulanda2024analysis}.

\subsection{Adaptive clustering}
\label{sec:pipeline}

Rather than applying a fixed partition of the PDT, we optimize the cluster
boundaries according to their weighted secret-key contribution. For each
candidate cluster, both CMA and de-fading are evaluated, such that the optimal
cluster size depends on the local channel statistics and the treatment of
residual fading. The adaptive clustering proceeds as follows:

\begin{enumerate}
    \item \textbf{Framing}: Measurements are divided into frames of shorter length, within each of which the transmittance can be  stationary.  Frames are defined by temporal order of measurements (unlike bins or clusters which group measurements purely by transmittance value) bounded from above by the temporal coherence of the channel and from below by the statistics required for parameter estimation. The parameters of the frames (noise $\varepsilon_j$ or transmittance $\eta_j$) can be estimated either by disclosing a fraction of the data or by performing information reconciliation prior to parameter estimation \cite{leverrier2015composable, jain2022practical}.
    \item \textbf{Threshold cluster}: Since higher mean transmittance tolerates larger \(\mathrm{Var}(\sqrt{\eta})\), we form an initial cluster starting from the frame with the highest transmittance and progressively add frames in order of decreasing transmittance. The cluster is chosen according to the maximum weighted SKR. When comparing weighted key rates, we evaluate the keys obtained using CMA and de-fading and choose the higher rate.
    \item \textbf{K-means pre-clustering}: The remaining data are then pre-grouped using $k$-means clustering to identify regions of the PDT with reduced internal transmittance fluctuations. The value of $k$ is chosen to maximize the number of clusters with a positive weighted SKR, identifying the regions of the PDT with the most favorable performance.
    \item \textbf{Cluster widening}: Beginning with the highest mean-transmittance cluster identified in step 3, the transmittance bounds of each cluster are gradually expanded to include neighboring frames (at the edges of the cluster) until the weighted SKR for that cluster is maximized.
    \item \textbf{Iteration over remaining clusters}: Step 4 is repeated for each remaining cluster from step 3 that has not already been absorbed by a previously expanded cluster, continuing until the full PDT is covered.
\end{enumerate}
The resulting partition therefore adapts both to the measured PDT and to the
residual-fading treatment, while maximizing the weighted SKR and retaining a
large fraction of the acquired data. 

\begin{figure}[!b]
    \centering
    \includegraphics[width=0.75\linewidth]{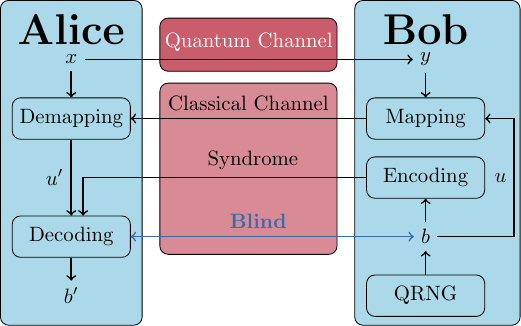}
    \caption{Rate-adaptive multidimensional reverse reconciliation. 
Alice and Bob obtain the correlated variables $x$ and $y$ from the quantum 
channel, respectively. Bob generates the binary sequence $b$ using a quantum 
random number generator (QRNG) and communicates the mapping and syndrome 
information over the authenticated classical channel. Alice recovers an 
estimate $b'$ through demapping and MET-LDPC decoding, with blind 
reconciliation providing additional information when required. }
    \label{fig:mdIR}
\end{figure}

\subsection{Rate-adaptive blind information reconciliation}
\label{sec:reconciliation}

Key extraction requires information reconciliation of Alice's and Bob's
correlated Gaussian variables. We implement multidimensional reverse
reconciliation~\cite{PhysRevA.77.042325}, which maps the Gaussian channel onto
a virtual binary-input channel and thereby supports operation at the low SNRs
encountered here, using multi-edge-type low-density parity-check (MET-LDPC)
codes~\cite{PhysRevA.103.062419}.

Although clustering reduces the variation of the channel parameters within
each cluster, the reconciliation frames it contains can still exhibit
different SNRs. A fixed-rate code must then either operate conservatively,
reducing the reconciliation efficiency for favourable frames, or incur a high
frame-error rate (FER) when the instantaneous SNR falls. This is the standard approach and we refer to it as the "one-shot" approach. Accounting for
decoding failures, the extractable rate is (\cite{laudenbach2018continuous})
\begin{equation}
R^{\mathrm{IR}}
=
\max\left\{
0,
(1-\mathrm{FER})
\left[
\beta I(A:B)^{x,p}-\chi_E
\right]
\right\},
\label{eq:IR-skr}
\end{equation}
where $\mathrm{FER}$ is the frame-error rate.

To accommodate the residual SNR variation, we employ rate-adaptive blind
reconciliation~\cite{DBLP:journals/qic/Martinez-MateoEM12}, illustrated in
Fig.~\ref{fig:mdIR}. For each reconciliation frame, the code is first punctured
to a rate set by the SNR estimate for that frame, and Alice attempts to
decode. If decoding fails, Bob progressively reveals punctured bits over the
authenticated classical channel, converting them into shortened bits and
thereby lowering the effective code rate. Alice repeats the decoding with the
additional information until it succeeds, or until further disclosure would no
longer yield a positive secret-key contribution.

The reconciliation rate therefore follows the SNR of each frame rather than
being fixed for the complete data set: frames that fail at the initial rate
can be recovered at a lower effective rate instead of being discarded, while
favourable frames retain a higher reconciliation efficiency. We use MET-LDPC
base codes of rate 0.005, 0.01, 0.02, and 0.05 with code lengths of 512000, 819200, 1024000, and 819200 respective, adapting their effective rates through
puncturing and shortening. Together with transmittance-based clustering and the treatment of residual fading, rate-adaptive reconciliation completes the adaptive post-processing framework used below for secret-key extraction.

\begin{figure*}[!b]
    \centering
    \includegraphics[width=1\linewidth]{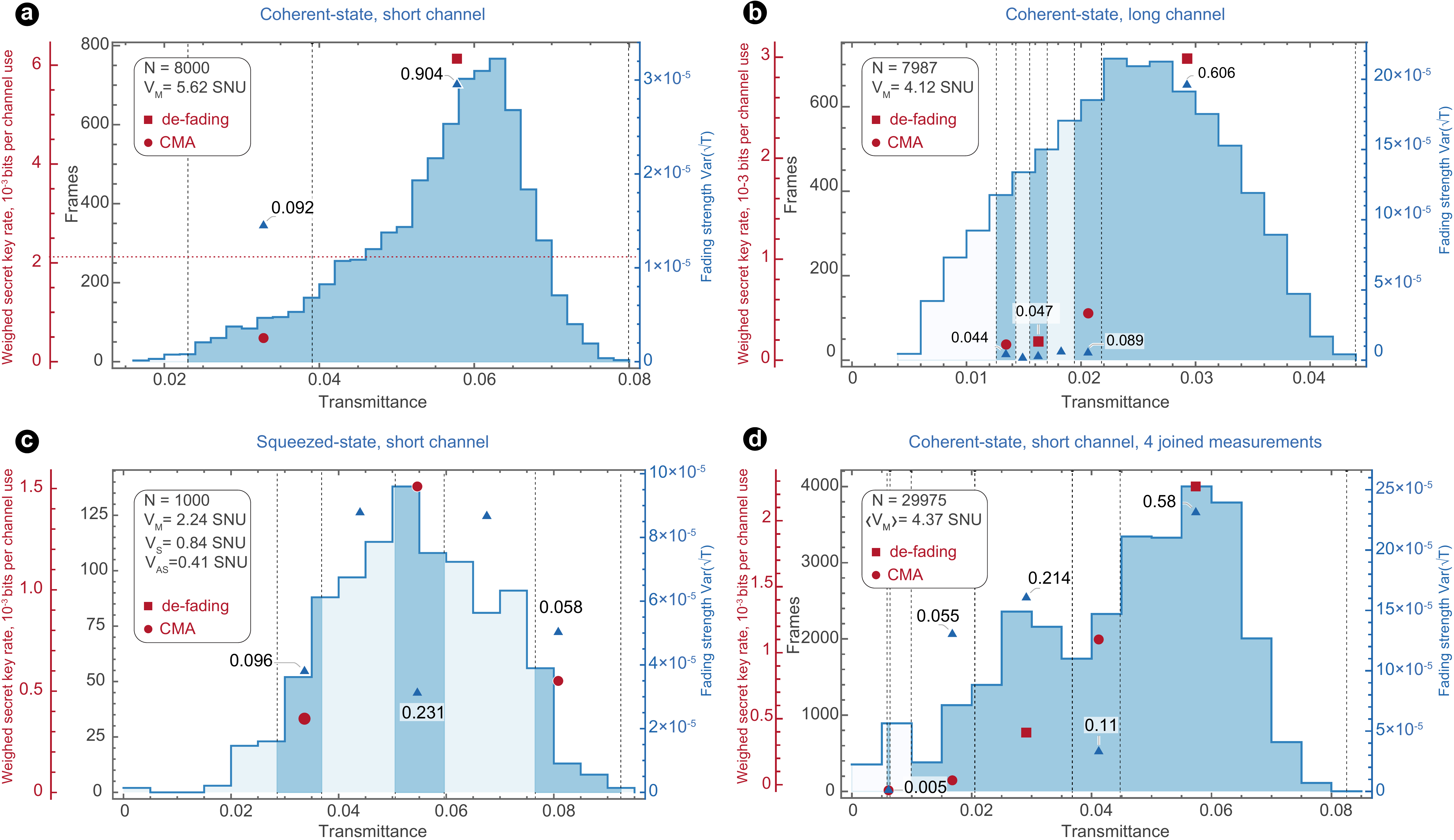}
    \caption{
Adaptive post-processing of measured probability distributions of transmittance (PDTs) under different hybrid-channel conditions. \textbf{a}, \textbf{b} and \textbf{d} show coherent-state measurements, and \textbf{c} shows a squeezed-state measurement. The blue histograms show the measured PDTs, with the number of frames given by the left black axis. Vertical dashed lines indicate the boundaries of the clusters selected by the adaptive procedure. For each cluster, the blue triangles show the fading strength $\mathrm{Var}(\sqrt{\eta})$ (right blue axis), while the red markers show the corresponding weighted asymptotic secret-key rate (upper/left red axis).
Red circles and squares denote clusters processed using covariance-matrix averaging (CMA) and de-fading, respectively. The horizontal position of each marker corresponds to the mean transmittance $\langle \eta\rangle$ of the cluster, and the numerical labels indicate the fraction of the total measurement contained within that cluster. The PDTs are constructed from frames of $1.25\times10^{4}$ symbols for the coherent-state measurements and
$1.25\times10^{5}$ symbols for the squeezed-state measurement. In \textbf{a}, the red dotted horizontal line indicates the weighted secret-key rate obtained by processing the complete PDT without clustering. Parameters of each data set are provided on each sub-plot: $N$ - total number of frames,  $V_M$ modulation variance, or mean modulation variance $\langle V_M\rangle$, $V_S$ - squeezing level and additional anti-squeezing $V_{AS}$.The detector efficiencies are 0.68 and 0.54 for the coherent- and squeezed-state protocols, respectively, and a reconciliation efficiency of $\beta=95\%$ is assumed. 
} 
    \label{fig:results-4}
\end{figure*}










\begin{table*}[t]
\centering
\small
\setlength{\tabcolsep}{3.2pt}
\setlength{\arrayrulewidth}{0.4pt}

\begin{tabular}{l|l|l|ccc|c|c}
\toprule
Data set &
\multicolumn{1}{c|}{%
\shortstack{QKD acquisition\\(CEST)}} &
\multicolumn{1}{c|}{%
\shortstack{Atmospheric\\averaging window}} &
\shortstack{Temperature\\{[$^\circ$C]}} &
\shortstack{Relative humidity\\{[\%]}} &
\shortstack{Wind speed\\{[km/h]}} &
\shortstack{Free-space loss\\{[dB]}} &
\shortstack{$D/r_0$} \\
\midrule

Fig.~4a &
28 Jul.~2025, 17:00 &
16:45--17:15 &
$23.8 \pm 0.1$ &
$64.1 \pm 0.6$ &
$4.8 \pm 2.5$ &
$8.0 \pm 0.8$ &
$1.75 \pm 0.27$ \\

Fig.~4b &
24 Jul.~2025, 16:33 &
16:18--16:48 &
$30.3 \pm 0.6$ &
$52.1 \pm 4.9$ &
$5.4 \pm 3.5$ &
$9.9 \pm 1.1$ &
$2.24 \pm 0.10$ \\

Fig.~4c &
31 Jul.~2025, 19:00 &
18:45--19:15 &
$28.3 \pm 0.2$ &
$56.9 \pm 1.0$ &
$6.8 \pm 3.0$ &
$12.6 \pm 2.5$ &
$1.98 \pm 0.28$ \\

Fig.~4d &
28 Jul.~2025, 16:00--17:18 &
16:00--17:18 &
$24.0 \pm 0.2$ &
$63.1 \pm 1.3$ &
$5.2 \pm 3.0$ &
$7.9 \pm 0.8$ &
$1.73 \pm 0.15$ \\

\bottomrule
\end{tabular}

\caption{Atmospheric conditions, free-space channel losses, and turbulence
strength corresponding to the measurements shown in
Fig.~\ref{fig:results-4}. The QKD acquisition times and analysis windows are
indicated for each data set. Atmospheric parameters were sampled once per
minute, whereas the channel
monitoring power used to determine the free-space
loss and $D/r_0$ was sampled at 10~Hz. All quantities are evaluated over the indicated intervals and reported as mean values and corresponding
standard deviations. For $D/r_0$, the reported statistics are calculated from
consecutive approximately 10-minute blocks. All times are expressed in Central European Summer Time (CEST).}
\label{tab:atmospheric_conditions}

\end{table*}

\section{Results}\label{sec:results}

We evaluate the adaptive post-processing framework on measurements acquired
under different atmospheric conditions, channel configurations and modulation
settings, summarised in Table~\ref{tab:atmospheric_conditions}.
Figure~\ref{fig:results-4} shows four representative data sets: coherent-state
measurements over the short (\textbf{a}) and long (\textbf{b}) channel, a
squeezed-state measurement over the short channel (\textbf{c}), and four
coherent-state measurements over the short channel processed jointly
(\textbf{d}). Together these span the range of mean transmittance and
fluctuation strength encountered on the deployed link. 

For the measurements in Fig.~\ref{fig:results-4}a--c the QKD acquisition is
considerably shorter than the window over which the atmospheric conditions
were averaged, and the two intervals are listed separately in
Table~\ref{tab:atmospheric_conditions}. Figure~\ref{fig:results-4}d instead
covers the entire acquisition period, during which the atmospheric conditions were
 continuously characterized.

\begin{figure}[!t]
    \centering
    \includegraphics[width=0.99\linewidth]{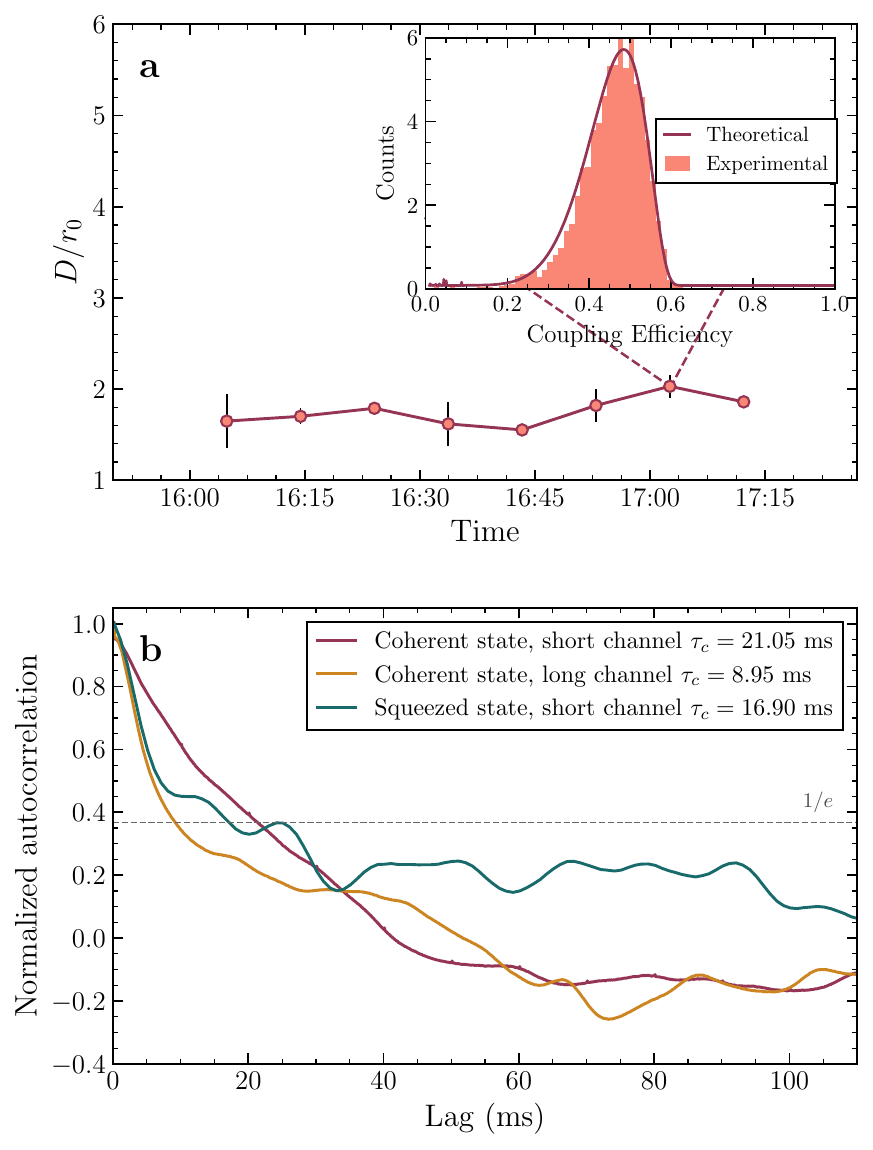}
    \caption{Characterisation of the free-space channel.
\textbf{a}, Turbulence strength, quantified by the ratio $D/r_0$ extracted in 10-minute intervals during
the acquisition of 28 July 2025 from the power coupled into single-mode fibre
with active tip/tilt correction. Error bars give the uncertainty of the fit.
Inset: measured distribution of the instantaneous coupling efficiency (bars)
and the fitted theoretical single-mode coupling-efficiency probability density function (line)
from which $D/r_0$ is obtained.
\textbf{b}, Normalised autocorrelation of the channel transmittance for three
representative measurements, computed from the frame-wise transmittance
estimates. The dashed line marks $1/e$, and the coherence time $\tau_c$ given
in the legend is the lag at which each autocorrelation crosses it.}
    \label{fig:turbulence}
    \label{fig:acf}
\end{figure}

\subsection{Channel dynamics and frame selection}
\label{sec:dynamics}

The transmittance fluctuations of the hybrid channel originate in the
free-space segment, where atmospheric turbulence distorts the wavefront and
thereby modulates the efficiency with which the received field couples into
single-mode fibre. We characterize the turbulence strength experienced by the free-space terminals through the ratio $D/r_0$, where $r_0$ is the Fried parameter expressed at 1546.92~nm \cite{fried1965statistics}. Although typically the diameter $D$ is referred to the diameter of the stop aperture at the receiver side, which is equivalent to $D = D_{Rx} = 0.4$~m, in this setup $D$ represents the diameter of the beam at the receiver's aperture, which due to the divergence imposed at the Tx, and its free-space propagation through 620~m, is completely collected by the Rx without clipping. To estimate this ratio, we apply the model of the coupling efficiency probability density function~\cite{Canuet2018}, with the modification regarding the finite-size of the control system proposed in Ref.~\cite{Roddier_1999}, to the measured coupled power in 10-minute
intervals. The $D/r_0$ parameter estimated for the time series of the acquisition listed in the last row of Tab.~\ref{tab:atmospheric_conditions}
is shown in Fig.~\ref{fig:turbulence}a,
together with a representative fit of the coupling efficiency (inset). Across the measurements of this work $D/r_0$ ranges
between $1.7$ and $2.2$ (Table~\ref{tab:atmospheric_conditions}) corresponding to a weak-turbulence regime \cite{tyson2022principles}, in which the wavefront error can be simply corrected by reducing tip-tilt fluctuations.
Thanks to the correction provided by the fast-steering mirror at Rx, the residual coupling efficiency
(Fig.~\ref{fig:turbulence}a, inset) is centered at $\approx0.48$ and spans
$\approx0.25$ to $0.55$, a range of about $3.4$~dB. It is this distribution
that the free-space segment imprints on the channel transmittance.

Turbulence of this strength fixes how widely the transmittance fluctuates;
how rapidly it does so determines the processing-frame length available to the
adaptive clustering. Since clustering requires the transmittance to be
approximately constant within each frame, the channel coherence time sets an
upper bound on the frame length, which we now establish.

In Fig.~\ref{fig:results-4}a--c, each measurement comprises $\approx10^{8}$
symbols, divided into frames of $1.25\times10^{4}$ symbols ($100~\mu$s) for
the coherent-state protocol and $1.25\times10^{5}$ symbols ($1$~ms) for the
squeezed-state protocol. Assuming that information reconciliation precedes
parameter estimation~\cite{leverrier2015composable,jain2022practical}, all
symbols within a frame contribute to the estimation of the channel parameters
rather than a disclosed subset. On this basis we evaluate $\eta_i$ and
$\varepsilon_i$ for every frame.
 
Figure~\ref{fig:acf} shows the normalised autocorrelation of the transmittance
for three representative measurements. We define the channel coherence time
$\tau_c$ as the lag at which the autocorrelation falls to $1/e$, the
convention that returns the decay constant directly for a first-order
stationary process with an exponential autocorrelation, as is commonly assumed
for atmospheric fading and slow drift of the fibre coupling. On this measure
we obtain $\tau_c=21.1$~ms for the coherent-state short-channel measurement,
$8.9$~ms for the coherent-state long-channel measurement, and $16.9$~ms for
the squeezed-state measurement, so that the coherence time lies between $9$
and $21$~ms across the conditions encountered. The measured autocorrelations
are not strictly exponential --- the squeezed-state trace retains a residual
correlation of $\approx0.2$ out to the longest lags, and the coherent-state
traces become weakly anticorrelated beyond $\approx50$~ms, indicating a
quasi-periodic contribution --- so $\tau_c$ is used here as an operational
measure of decorrelation rather than as a fitted model parameter.
 
The three measurements were acquired on different days and therefore under
different atmospheric conditions
(Table~\ref{tab:atmospheric_conditions}). The variation in $\tau_c$ reflects
these conditions rather than the channel configuration, since the free-space
segment is common to both configurations and the additional fibre span of the
long channel is passive, and illustrates the range of channel dynamics the
processing must accommodate.
 
The frame durations used above are accordingly short compared with the
measured coherence times, by a factor of $90$ to $210$ for the coherent-state
protocol and by a factor of $17$ for the squeezed-state protocol. The
transmittance can therefore be treated as constant within a frame for both
protocols, and the stationarity bound is not the binding constraint: the frame
length is set in practice by the statistics required for parameter estimation.
The longer squeezed-state frames consequently resolve the transmittance
distribution more coarsely, which constrains the formation of low-fluctuation
clusters. The frame-wise estimates $\eta_i$ then define the PDTs shown in
Fig.~\ref{fig:results-4}.


\subsection{Secret-key extraction across different fading regimes}

For each measurement we compare three treatments of the transmittance
distribution: processing the complete PDT without post-selection, conventional
threshold post-selection and uniform binning, and the adaptive clustering of
Sec.~\ref{sec:pipeline}. Within each selected data region the residual fading
is treated using either CMA or de-fading, whichever yields the larger weighted
secret-key contribution. The total secret-key rate therefore accounts both for
the rate obtained within each cluster and for the fraction of the acquired
data retained for key extraction.

Under comparatively favorable channel conditions
(Fig.~\ref{fig:results-4}a), the complete PDT yields a positive asymptotic
secret-key rate even without post-selection, as indicated by the red dotted
line. Nevertheless, processing the complete distribution together averages
measurements acquired at different transmittances and therefore retains the
full fading penalty. Dividing the PDT into optimized clusters reduces
$\mathrm{Var}(\sqrt{\eta})$ within the individual data regions and increases the total weighted secret-key rate (red points in Fig.~\ref{fig:results-4}a). In this measurement, less than 1\% of the
highest-loss data needs to be discarded, showing that clustering can improve the key rate even when almost the complete measurement remains usable. 

Figure~\ref{fig:results-4}a also illustrates the complementary roles of CMA and de-fading approaches.  CMA is advantageous for narrow clusters with small residual fluctuations because it preserves the measured signal level while incorporating fading into the security bound. For broader clusters, the fading-induced penalty can instead outweigh the additional
attenuation introduced by de-fading, making de-fading favorable.  The optimal treatment of a cluster therefore depends not only on its mean transmittance but also on the residual fluctuations within it. This complementarity is incorporated directly into the cluster optimization.

As the channel conditions become less favorable (mean transmittance becomes lower), processing the complete PDT no longer yields a positive secret key and clustering becomes necessary. For the short channel, the threshold cluster alone retains, on average, 85\% of the data and  contributes 91\% of the total SKR, so the remaining clusters contribute only marginally. For the long channel, at an intermediate mean transmittance, the threshold cluster still yields the higher key rate, providing on average 70\% of the total SKR, but now captures only 42\% of the data. The additional clusters become more important, capturing 29\% more data and 30\% of the total SKR on average. While the averaged statistics discussed here and further are not shown explicitly, they are computed from the fully processed results for all measurements, provided in a dedicated repository \cite{zdataset2026}. Higher mean transmittance is generally accompanied by broader PDTs and, correspondingly, broader optimal clusters. The example shown in Fig.~\ref{fig:results-4}b lies toward the favorable end of this regime rather than representing a typical instance, so the relative contribution of the additional clusters can be considerably more pronounced for less favorable realizations at comparably low mean transmittance. \\
As the mean transmittance decreases further, the optimal clusters narrow, favoring CMA over de-fading. The threshold cluster remains relevant, but most of the key now originates from the additional clusters. In this low-transmittance regime, the key rate becomes especially sensitive to the estimated excess noise variance, so each cluster must contain enough data for a precise estimate. This requirement becomes even more important in the finite-size regime, where limited statistics are the main obstacle to positive key generation \cite{pirandola2021composable}.

The same underlying adaptive-processing principle also applies to the squeezed-state protocol, as shown in
Fig.~\ref{fig:results-4}c. Three key-generating clusters are identified for
this measurement, demonstrating that transmittance-based clustering and
residual-fading mitigation remain applicable when the underlying quantum
state and security analysis are changed. The squeezed-state protocol,
however, additionally requires characterization of the squeezing variance
$V_S$ and the anti-squeezing noise $V_{AS}$. Accurate estimation of these
parameters requires larger processing frames than for the coherent-state
measurements, here $1.25\times10^{5}$ compared with
$1.25\times10^{4}$ symbols. This reduces the resolution with which the
transmittance distribution can be partitioned and places a 
statistical constraint on the formation of low-fluctuation clusters.


 \begin{figure}[!t]
    \centering
    \includegraphics[width=0.99\linewidth]{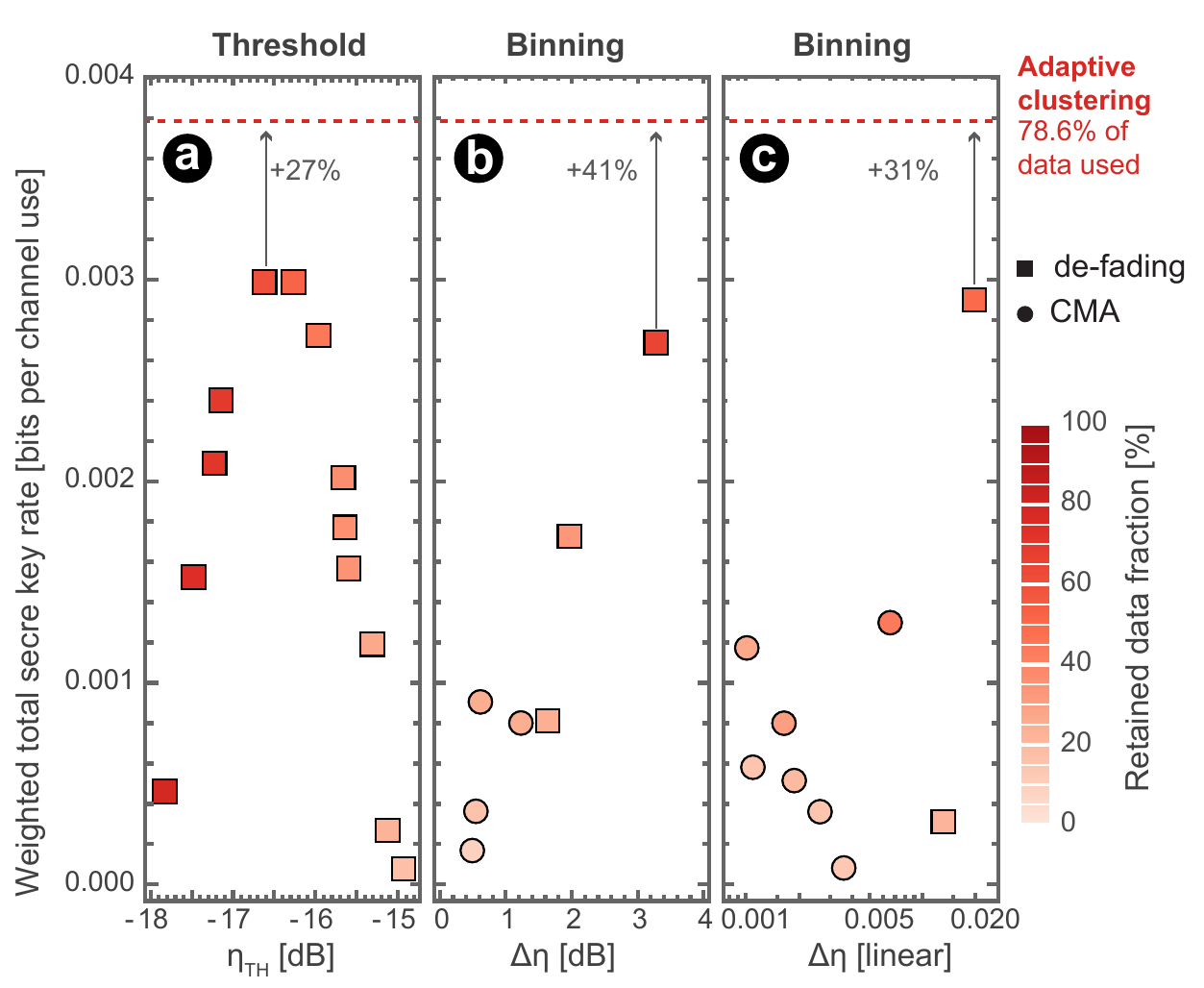}
    \caption{Comparison of weighted total secret key rates for different post-selection approaches, for the same data set shown in Fig.~\ref{fig:results-4}b: (a) threshold post-selection, (b) uniform binning of the PDT in logarithmic scale, with $\Delta\eta$ width of the bin width, and (c) uniform binning in linear scale. Point color indicates the fraction of the data retained for key extraction. The dashed red line indicates the key rate obtained using adaptive clustering.  Arrows show the relative improvement in total key rate over the optimum of each individual post-selection approach.}
    \label{fig:compare}
\end{figure}

A more detailed comparison of post-selection approaches for the data set of Fig.~\ref{fig:results-4}b is shown in Fig.~\ref{fig:compare}. For threshold post-selection (Fig.~\ref{fig:compare}a), lowering $\eta_{\mathrm{th}}$ from $-14$ to $-18$~dB progressively captures more data thus reducing the weighting penalty $p(\eta_{\mathrm{th}})$ on the SKR, but at the cost of admitting more fading (i.e. a increasing $\mathrm{Var}(\sqrt{\eta})$, which is relevant for CMA method) and decreasing the SNR (which is relevant for de-fading method). The balance between these effects is intricate, depending on channel parameters such as the excess noise and on the shape of the PDT. \\
Uniform binning can be performed either on a linear transmittance scale or on a dB scale, the latter corresponding to wider bins at higher mean transmittance. Figure~\ref{fig:compare}b,c, show broader bins favour de-fading while narrower bins favour CMA.

On this data set, adaptive clustering exceeds the optimum of every
conventional strategy: by $27\%$ over the best threshold, by $31\%$ over the best linear binning and by $41\%$ over the best dB binning, while retaining $78.6\%$ of the acquired data against $60.5,\, 65.0,\, 49.6\%$ for the respective optima. The same holds across all measurements, for both protocols and both channel configurations~\cite{zdataset2026}. The optimal bin width is moreover not transferable between measurements: no single linear-scale width is consistently preferred, whereas on the dB scale $\Delta\eta$ widths of 1--1.5~dB are favoured on average in this regime. Adaptive clustering requires no such choice, since the cluster boundaries follow from the measured key-rate performance. A finely tuned bin width, which further depends on the minimum bin population imposed, can occasionally exceed the key rate of adaptive clustering, but only by concentrating on a single narrow high-transmittance bin while discarding nearly all remaining data, which is impractical and unsuitable for the finite-size or the composable regimes.


%
%

\begin{table*}[!t]
\centering

\begin{subtable}{\textwidth}
\centering
\caption{Coherent-state protocol over the short channel.}
\label{tab:IRresults2}
\vspace{0.3em}
\begin{tabular}{l|rr|rrr|rrr|rr}
\toprule
& \multicolumn{2}{c|}{Channel parameters}
& \multicolumn{3}{c|}{One-shot}
& \multicolumn{3}{c|}{Blind}
& \multicolumn{2}{c}{Increase}\\
Cluster & \multicolumn{1}{c}{Mean loss (dB)} & \multicolumn{1}{c|}{$\varepsilon$ ($\mu$SNU)}
& SKR & FER & Length & SKR & FER & Length & SKR & Length\\
\midrule
1 & \multicolumn{1}{c}{12.2} & \multicolumn{1}{c|}{1730}
  & $2.31\times10^{-3}$ & 0.69 & 221341
  & $2.69\times10^{-3}$ & 0.60 & 257526 & 16.35\% & 16.35\%\\
2 & \multicolumn{1}{c}{13.8} & \multicolumn{1}{c|}{1.87}
  & $4.88\times10^{-4}$ & 0.40 & 46783
  & $5.75\times10^{-4}$ & 0.37 & 54446 & 17.94\% & 16.38\%\\
3 & \multicolumn{1}{c}{14.9} & \multicolumn{1}{c|}{1490}
  & $5.89\times10^{-5}$ & 0.43 & 5162
  & $7.29\times10^{-5}$ & 0.14 & 6394 & 23.85\% & 23.85\%\\
\midrule
Total & \multicolumn{1}{c}{--} & \multicolumn{1}{c|}{--}
  & $2.86\times10^{-3}$ & -- & 273286
  & $3.34\times10^{-3}$ & -- & 318366 & 16.77\% & 16.50\%\\
\bottomrule
\end{tabular}
\end{subtable}

\vspace{1em}

\begin{subtable}{\textwidth}
\centering
\caption{Coherent-state protocol over the long channel.}
\label{tab:IRresults1}
\vspace{0.3em}
\begin{tabular}{l|rr|rrr|rrr|rr}
\toprule
& \multicolumn{2}{c|}{Channel parameters}
& \multicolumn{3}{c|}{One-shot}
& \multicolumn{3}{c|}{Blind}
& \multicolumn{2}{c}{Increase}\\
Cluster & \multicolumn{1}{c}{Mean loss (dB)} & \multicolumn{1}{c|}{$\varepsilon$ ($\mu$SNU)}
& SKR & FER & Length & SKR & FER & Length & SKR & Length\\
\midrule
1 & \multicolumn{1}{c}{15.2} & \multicolumn{1}{c|}{0.47}
  & $2.40\times10^{-3}$ & 0.50 & 193211
  & $2.66\times10^{-3}$ & 0.44 & 213584 & 11.08\% & 10.54\%\\
2 & \multicolumn{1}{c}{16.7} & \multicolumn{1}{c|}{465}
  & $2.16\times10^{-4}$ & 0.46 & 17333
  & $2.62\times10^{-4}$ & 0.44 & 20961 & 21.19\% & 20.93\%\\
3 & \multicolumn{1}{c}{17.4} & \multicolumn{1}{c|}{61.5}
  & $6.79\times10^{-5}$ & 0.00 & 5060
  & $6.86\times10^{-5}$ & 0.00 & 5057 & 1.05\% & $-0.05$\%\\
4 & \multicolumn{1}{c}{17.9} & \multicolumn{1}{c|}{43.3}
  & $1.01\times10^{-4}$ & 0.07 & 7960
  & $1.16\times10^{-4}$ & 0.07 & 9068 & 15.23\% & 13.92\%\\
5 & \multicolumn{1}{c}{18.4} & \multicolumn{1}{c|}{52.5}
  & $8.28\times10^{-5}$ & 0.07 & 6211
  & $1.01\times10^{-4}$ & 0.07 & 7957 & 22.02\% & 28.10\%\\
\midrule
Total & \multicolumn{1}{c}{--} & \multicolumn{1}{c|}{--}
  & $2.86\times10^{-3}$ & -- & 229775
  & $3.21\times10^{-3}$ & -- & 256627 & 12.06\% & 11.68\%\\
\bottomrule
\end{tabular}
\end{subtable}

\vspace{1em}

\begin{subtable}{\textwidth}
\centering
\caption{Squeezed-state protocol over the short channel.}
\label{tab:IRresultsSQZ}
\vspace{0.3em}
\begin{tabular}{l|rr|rrr|rrr|rr}
\toprule
& \multicolumn{2}{c|}{Channel parameters}
& \multicolumn{3}{c|}{One-shot}
& \multicolumn{3}{c|}{Blind}
& \multicolumn{2}{c}{Increase}\\
Cluster & \multicolumn{1}{c}{Mean loss (dB)} & \multicolumn{1}{c|}{$\varepsilon$ ($\mu$SNU)}
& SKR & FER & Length & SKR & FER & Length & SKR & Length\\
\midrule
1 & \multicolumn{1}{c}{10.9} & \multicolumn{1}{c|}{368}
  & $1.80\times10^{-3}$ & 0.00 & 41276
  & $1.89\times10^{-3}$ & 0.00 & 42359 & 5.18\% & 2.63\%\\
2 & \multicolumn{1}{c}{12.6} & \multicolumn{1}{c|}{90.0}
  & $3.60\times10^{-3}$ & 0.15 & 86226
  & $4.56\times10^{-3}$ & 0.04 & 108350 & 26.59\% & 25.66\%\\
3 & \multicolumn{1}{c}{14.7} & \multicolumn{1}{c|}{142}
  & $8.96\times10^{-4}$ & 0.19 & 20802
  & $1.03\times10^{-3}$ & 0.04 & 23881 & 14.80\% & 14.80\%\\
\midrule
Total & \multicolumn{1}{c}{--} & \multicolumn{1}{c|}{--}
  & $6.30\times10^{-3}$ & -- & 148304
  & $7.48\times10^{-3}$ & -- & 174590 & 18.79\% & 17.72\%\\
\bottomrule
\end{tabular}
\end{subtable}

\caption{Weighted secret-key rates (SKR, bits per channel use) and key lengths
(bits) obtained with one-shot and blind information reconciliation for the
coherent- and squeezed-state measurements. Excess noise $\varepsilon$ is
referred to the channel output and expressed in units of $10^{-6}$ shot-noise
units ($\mu$SNU).}
\label{Tab:IR}
\end{table*}

\subsection{Rate-adaptive information reconciliation}

We next evaluate whether the adaptation to the fluctuating channel can be
extended from the security analysis to practical information reconciliation. We apply the reconciliation procedure described in
Sec.4~\ref{sec:reconciliation} to the measurements in
Fig.~\ref{fig:results-4}a--c processed using CMA, and compare rate-adaptive
blind reconciliation with one-shot decoding, in which no additional
information is revealed after an unsuccessful decoding attempt. The results with the channel parameters of each cluster
are summarized in Table~\ref{Tab:IR}. The secret-key rate (SKR) reported is obtained after the information reconciliation based on Equation~\ref{eq:IR-skr}. For the one-shot results $\beta I$ is replaced by the code rate. For the blind results this applies on a per frame basis and reported number is a weighted average. 

Across the three measurements, blind reconciliation increases the total
secret-key rate by 12.06\% for the coherent-state long-channel measurement,
16.77\% for the coherent-state short-channel measurement, and 18.79\% for
the squeezed-state measurement. The corresponding extracted key lengths
increase by 11.68\%, 16.50\%, and 17.72\%, respectively. The improvement is
therefore not restricted to a particular channel configuration or quantum
state.

Information reconciliation can improve the SKR in two ways, as evident from Equation~\ref{eq:IR-skr}: by decreasing the FER, or by increasing $\beta$. The gain from blind reconciliation depends on the performance of the one-shot decoder—frames with a high one-shot FER benefit substantially from adaptive rate matching. For example, in cluster~3 of the short-channel measurement (Table~\ref{tab:IRresults2}), the FER decreases from 0.43 to 0.14, increasing the secret-key rate by 23.85\%. Similarly, for cluster~2 of the squeezed-state measurement (Table~\ref{tab:IRresultsSQZ}), the FER decreases from 0.15 to 0.04, increasing the SKR by 26.59\%. By contrast, when the one-shot FER is already close to zero, as in cluster~3 of the long-channel measurement, blind and one-shot reconciliation yield nearly identical performance.

Clusters~4 and~5 of the long-channel measurement (Table~\ref{tab:IRresults1}) illustrate a complementary mechanism: the FER remains constant, yet due to a higher average code rate, the SKR improves by 15.23\% and 22.02\%, respectively. Here, blind reconciliation lets us select a higher code rate, recovering more bits from favorable frames, while shortening still allows less favorable frames to be decoded successfully. 

These results show that the principal benefit of blind reconciliation is the
recovery of frames that would otherwise be discarded because of a mismatch
between the instantaneous SNR and the initially selected code rate. In addition, blind reconciliation allows us to set the code rate closer to the capacity of the channel, thereby recovering more bits per frame without the association penalty to the FER. The
adaptation introduced at the level of the transmittance distribution is
therefore complemented by rate adaptation at the reconciliation-frame level.
Using the complete adaptive processing chain, we extract up to 318~kbit of
secret key from a single measurement over the fluctuating hybrid channel in the asymptotic regime

\subsection{Statistical requirements for finite-size composable security}

The number of symbols available within each cluster becomes increasingly
important when moving from asymptotic to finite-size and composable security.
One approach to increasing the available statistics is to jointly process
measurements acquired under comparable channel configurations, analogous to combining multiple satellite passes in satellite-based DV-QKD \cite{sidhu2023finite,marulanda2024analysis,brooke2026decoy}. 

Figure~\ref{fig:results-4}d shows the PDT obtained by combining four
coherent-state measurements over the short hybrid channel, acquired at
different times and with different modulation variances. Over the combined
acquisition period the mean free-space loss remained stable at
$7.9\pm0.8$~dB (Table~\ref{tab:atmospheric_conditions}), while individual
fades reached $\approx20$~dB. The PDT therefore broadens through fast
transmittance fluctuations and through the differing modulation variances of
the individual acquisitions, rather than through slow drift of the channel. Joint processing increases the number of symbols available for parameter
estimation within the selected clusters, but does not necessarily increase
the asymptotic secret-key rate because the combined data also contain the
transmittance fluctuations of the individual acquisitions and variations in
the modulation variance. Nevertheless, the increased block size reduces the
statistical uncertainty in the estimated channel parameters and is therefore important for approaching composable finite-size operation. The
highest-transmittance cluster in Fig.~\ref{fig:results-4}d contains
$2.15\times10^{8}$ symbols. Under the present experimental parameters, this
block size remains insufficient for positive composable key extraction, even with $\epsilon$-optimization~\cite{mountogiannakis2026optimizing} and a high information reconciliation efficiency.


\section{Discussion}

Our results demonstrate that a common CV-QKD architecture can operate across deployed fiber and free-space links despite their substantially different channel conditions. A key element enabling this interoperability is the ability of the post-processing to adapt to the measured channel
statistics without requiring a separate CV-QKD architecture for each
transmission regime. The hybrid link therefore illustrates a broader
principle for heterogeneous quantum networks: adaptation to the transmission medium can, to a substantial extent, be transferred from the quantum-optical layer to the subsequent processing of the measured data.

The three processing stages are intrinsically coupled through the channel statistics. A broad cluster retains more data but increases residual fading and broadens the SNR distribution presented to the reconciliation stage, whereas narrower clusters suppress these effects at the cost of fewer samples for parameter estimation. The comparison with threshold post-selection and uniform binning further shows that the SKR gain does not originate simply from partitioning the PDT, but from adapting the partition to the measured channel statistics. This trade-off becomes particularly important at low mean transmittance, where the key rate is increasingly sensitive to residual fading and parameter-estimation uncertainty. Conversely, retaining a broader range of channel realizations increases the SNR variation that reconciliation must accommodate, precisely the regime in which rate-adaptive decoding provides the largest benefit. The gains reported here therefore arise from treating clustering, residual-fading mitigation, and reconciliation as a coupled processing chain rather than as independent operations.

The 12--19\% improvement obtained with blind reconciliation illustrates this coupling particularly clearly. Even after clustering has reduced the
variation of the channel parameters, residual SNR variations remain between
reconciliation frames. Interactive rate adaptation allows frames that fail
at their initial code rate to be recovered at a lower effective rate rather
than being discarded. This improvement comes at the cost of additional
decoding attempts and classical communication, introducing a trade-off
between rate-adaptation granularity, information disclosure, and
computational complexity. Related interactive approaches, such as Multiple Decoding
Attempts~\cite{gumucs2021novel}, may further reduce the computational
cost by combining incremental information disclosure with early decoder
termination and an adaptive iteration budget. These considerations will
become increasingly important when moving towards real-time reconciliation
of fluctuating channels.


An important next step is to extend the complete framework to composable
finite-size security. Adaptive clustering introduces an inherent statistical
trade-off: narrowing a cluster reduces the residual fading penalty but also
reduces the number of symbols available for parameter estimation. In the
present measurements, combining several acquisitions increases the largest
cluster to $2.15\times10^{8}$ symbols, but this remains insufficient for
positive composable key extraction under the current system parameters.
Increasing the number of symbols acquired within a sufficiently stationary
channel interval is therefore important. Higher symbol rates enabled by
wider-bandwidth detectors and faster data converters, together with
real-time DSP and information reconciliation
\cite{hajomer2024continuous,hajomer2026chip}, would simultaneously increase
the available statistics and permit shorter processing intervals for
tracking faster channel variations. For the squeezed-state protocol, this
requirement is particularly relevant because characterization of the
squeezing and anti-squeezing parameters places additional demands on the
available statistics.

More broadly, the framework developed here is not specific to the
fiber--free-space channel of the present experiment. Its underlying
requirement is a channel whose transmittance can be estimated over intervals that are sufficiently stationary for parameter estimation and reconciliation. This situation arises naturally in terrestrial free-space links and is also expected in satellite quantum communication, where atmospheric propagation, pointing errors, and changing link geometry produce time-dependent transmittance, crucial for continuous-variable protocols \cite{Dequal2021,Derkach2020}. Receiver-generated local-oscillator CV-QKD combined with adaptive fading mitigation and rate-adaptive reconciliation therefore provides a route towards heterogeneous quantum networks in which fiber, terrestrial free-space, and future satellite links can be incorporated within a common CV quantum-communication architecture.

\FloatBarrier
\begin{backmatter}

\bmsection{Acknowledgments} 
 We acknowledge support from the European Union’s Horizon Europe research and innovation programme under the project “Quantum Secure Networks Partnership” (QSNP, Grant Agreement No. 101114043).  AAEH, HQN, ULA and TG acknowledge support from the Danish National Research Foundation through the Center for Macroscopic Quantum States (bigQ, DNRF142). AAEH, HQN, ULA, TG, SF and ABK acknowledge support from Innovation Fund Denmark through the CyberQ project (Grant Agreement No. 3200-00035B). AAEH, ULA and TG acknowledge support from Innovation Fund Denmark through the AccessQKD project (Grant Agreement No. 43560-00004B). RF acknowledges funding from the European Union's Horizon Europe
research and innovation programme under grant agreement No 101257526 (SUPERSPIN) and  the Quantera project CLUSSTAR (8C24003)
of MEYS, Czech Republic. Project CLUSSTAR has also received funding from the European Union’s Horizon
2020 Research and Innovation Programme under Grant
Agreements No. 731473 and No. 101017733 (QuantERA). RF and VU acknowledge the Project No. CZ.02.01.01/00/22\_008/0004649 (QUEENTEC) of MEYS. VU acknowledges support from the Czech Science Foundation under Project No. 21-44815L.


\bmsection{Data availability} Data underlying the results presented in this paper are available in \cite{zdataset2026}.

\smallskip

\bmsection{Disclosures} The authors declare no conflicts of interest.

\bmsection{Author contributions statement}
AAEH and HQN performed the experiments and the overall data processing. ID developed the post-processing framework and performed security modeling and analysis, with input from AAEH and under the supervision of VU. ABK developed and performed the information reconciliation under the supervision of SF. ER, MS, and YP established the hybrid testbed under the supervision of MA, GV, and PV. MH provided the LDPC code with a rate of 0.005. ER and MS characterized the free-space channel. AAEH wrote the manuscript with significant input from ID and ER and contributions from the other co-authors. ULA, PV, GV, and TG supervised the overall project. All authors contributed to the discussion and interpretation of the results.

\bigskip

\end{backmatter}

\bibliography{sample}

\bibliographyfullrefs{sample}


\ifthenelse{\equal{\journalref}{aop}}{%
\section*{Author Biographies}
\begingroup
\setlength\intextsep{0pt}
\begin{minipage}[t][6.3cm][t]{1.0\textwidth} 
  \begin{wrapfigure}{L}{0.25\textwidth}
    \includegraphics[width=0.25\textwidth]{john_smith.eps}
  \end{wrapfigure}
  \noindent
  {\bfseries John Smith} received his BSc (Mathematics) in 2000 from The University of Maryland. His research interests include lasers and optics.
\end{minipage}
\begin{minipage}{1.0\textwidth}
  \begin{wrapfigure}{L}{0.25\textwidth}
    \includegraphics[width=0.25\textwidth]{alice_smith.eps}
  \end{wrapfigure}
  \noindent
  {\bfseries Alice Smith} also received her BSc (Mathematics) in 2000 from The University of Maryland. Her research interests also include lasers and optics.
\end{minipage}
\endgroup
}{}

\end{document}